# Guiding the surgical gesture using an electro-tactile stimulus array on the tongue:

# A feasibility study

F. Robineau [*], F. Boy [**], J-P. Orliaguet [**], J. Demongeot [*] & Y. Payan [*].

[*] Technique in Imaging, Modeling and Cognition Laboratory (TIMC) – Institute for Applied Mathematics in Grenoble (IMAG) – CNRS – UMR 5525 - Faculty of Medicine, F-38706, La Tronche, Cedex, France.

[**] Perception-Action Team, Laboratory of Psychology and NeuroCognition (LPNC) CNRS - UMR 5105, Pierre Mendès-France University, F-38042 Grenoble Cedex 9, France.

Corresponding authors:    Yohan Payan
TIMC-IMAG – Equipe GMCAO
Pavillon Taillefer – Faculte de Medecine
38706 La Tronche - FRANCE
Yohan.Payan@imag.fr

**Abstract** — Under conventional "open-" surgery, the physician has to take care of the patient, interact with other clinicians and check several monitoring devices. Nowadays, the Computer Assisted Surgery proposes to integrate 3D cameras in the operating theatre in order to assist the surgeon in performing minimal-invasive surgical punctures. The cameras localize the needle and the computer guides the surgeon towards an intracorporeal clinically-defined target. A visualization system (screen) is employed to provide the surgeon with indirect visual spatial information about the intracorporeal positions of the needle. The present work proposes to use another sensory modality to guide the surgeon thus keeping the visual modality fully dedicated to the surgical gesture. For this, the sensory substitution paradigm using the Bach-y-Rita's "Tongue Display Unit" (TDU) is exploited to provide to the surgeon information of the position tool. The TDU device is composed of a 6×6 matrix of electrodes transmitting electrotactile information on the tongue surface. The underlying idea consists in transmitting information about the deviation of the needle movement with regard to a pre-planned "optimal" trajectory. We present an experiment assessing the guidance effectiveness of an intracorporeal puncture under TDU guidance with respect to the performance evidenced under a usual visual guidance system.

# 1 Introduction

## 1.1 Guidance of intracorporeal puncture

The development of Computer-Aided Surgical (CAS) systems to assist surgeons in the realization of minimally invasive (MI) intracorporeal gestures is a growing domain aiming at improving the safety and the efficiency of the therapeutical procedures [1, 2, 3]. More specially, the present research focuses on MI percutaneous gestures. Such procedures involve passing an ancillary (e.g., a screw or a biopsy needle) to a previously defined anatomical target inside the patient's body through small incisions made in the skin. The entire procedure is carried out by manipulating the external end of the ancillary from outside the patient's body, the surgeon being deprived of direct visual feedback about the actual ancillary trajectory. This is the case, for example, in iliosacral screw placement [4], percutaneous pericardiocentesis [5], renal puncture [6] or transhepatic cholangiography [7].

CAS systems propose several solutions to guide the surgeon's action towards the anatomical target. Prior to the therapeutical gesture, images of the patient's body (e.g., CT scan, MRI or Ultrasounds images) are collected to allow the planning of an "optimal" trajectory between an entry point located on the skin surface and the anatomical target. The conventional CAS procedure involves fastening Infra-Red (IR) positioning markers onto the ancillary (a needle in the case of a puncture) and placing IR cameras into the operating theatre in order to compute the position of the ancillary. The actual trajectory of the ancillary is therefore compared to the planned trajectory thus providing the surgeon with information about the deviation of his/her gesture with respect to the "optimal" planned trajectory. A "confidence tunnel" revolving around the planned trajectory is calculated, thus allowing the surgeon to master the unavoidable deviations to the optimal trajectory. Most conventional CAS systems use visualization devices (a screen) that provide the surgeon with indirect visual spatial information about the intracorporeal positions of the ancillary and thus allow the control of the puncturing needle.

However one limitation can be drawn from such systems using visual guidance: during movement guidance, the surgeon needs to look at the screen and therefore cannot focus his visual attention on the patient's body surface. This limitation is probably the most significant and is particularly crucial for procedures like Percutaneous Renal Puncture (PRP) or Percutaneous TransHepatic Cholangiography (PTHC). Indeed, in both cases, the surgeon has to gaze at the external tip of the needle once it reaches the target (the calices of the kidney for the PRP, the biliary radicles for the PTHC), in order to check for body fluid flowing out (either urine for PRP or bile for PTHC).

The present study aims at evaluating a novel and alternative answer to the challenges imposed by the MI guidance of an ancillary. The solution consists in presenting the feedback about the position of the surgical instrument with respect to the planned trajectory in another sensory modality, namely the lingual tactile modality. The underlying idea is to maintain the surgeon fully dedicated to the therapeutic procedure (vision focused on the insertion area, communication with the assistants…), and to use another unexploited sensory modality - the tactile modality - to convey information about the trajectory of the ancillary. This principle of "sensory substitution", or more generally speaking of "perceptual supplementation", was introduced and extensively studied by Paul Bach-Y-Rita and colleagues in the context of tactile visual substitution systems. These authors [8, 9] evidenced that stimulus characteristics of one sensory modality (e.g., a visual stimulus) could be transformed into stimulations of another sensory modality (e.g., a tactual stimulation). In the case of tactile visual substitution, the so-called "sensory substitution" thus involves the conversion of light energy into mechanical (or electrical) energy. In this case, on a perceptual point of view, this principle involves transcoding the visual percept into a tactual one.

## 1.2 Sensory substitution: the TDU system

The first development of tactile visual substitution systems (TVSS) was designed to provide distal spatial information to blind people [10, 11]. The original TVSS was composed of 400 stimulators (20×20 matrix, Ø 1mm each) placed on the chest or on the brow and rendered the images captured by a video camera into a "tactile image". Their results showed that blind subjects were capable to achieve 100% correct performance in an object recognition task after a short prior training (50 trials). A number of converging experimental observations concluded that, after adequate training, subjects, using a TVSS fulfilled a shape-recognition task and even experienced a "projection" of the objects they tactually perceived in the external world [12]. For instance, a sudden change in the zoom of the camera caused participants to act as if they were approaching an obstacle. This latter phenomenon was interpreted in terms of functional similarities between primary visual area (occipital cortex) and somatosensory areas (parietal cortex) that allow the reception and the processing of afferent signals originating in different sensory systems [13].

Tackling the practical problems posed by the mechanical stimulation of the skin, Bach-y-Rita and his collaborators converged towards the electro-stimulation of the tongue surface [14]. The human tongue is indeed a highly dense [15], sensitive and discriminative (spatial threshold = 2 mm) array of tactile receptors that are similar to the ones of the skin. Moreover, the high conductivity offered by the saliva insures a highly efficient electrical contact between the electrodes and the tongue surface. It is to note that these latter characteristics highlight that the acuity of the tongue is higher than the one of the tip of the index fingers, and that the receptors located on the surface can be easily and selectively activated. Therefore, Bach-y-Rita and colleagues designed and evaluated a practical human-machine interface, termed the Tongue Display Unit (TDU). The TDU [16] consists in a 2D array of miniature electrodes (12×12 matrix) held between the jaws and positioned in close contact with the anterior-superior surface of the tongue. The functioning of the TDU matrix is monitored by a computer-controlled external triggering device. Each electrode can be individually activated, thus stimulating a restricted area on the tongue. The amplitude and the intensity of the electrical pulses are adjusted to produce vibrational sensations on the tongue. Several experiments [e.g., see 16] showed that after adequate training with a TDU, blind subjects experienced the same "projection" of the objects as the one evidenced when using a classical TVSS system.

The present research aims at evaluating the feasibility of the TDU guidance in the context of MI surgical gesture. As a first evaluation, it was proposed to carry out experiments in order to quantitatively compare the TDU guidance with the usual visual guidance of a MI intracorporeal puncture.

## 2    Experiment

### 2.1.    The puncture task

Subjects were required to carry out punctures on an artificial model of the abdominal cavity (Phantom, CIRS®, Norfolk, VA, USA) placed in ventral decubitus. The experimental task was defined as follows: Participants had to progressively introduce a surgical needle inside the Phantom (L = 12cm, Ø = 1mm) until reaching an intracorporeal spherical target (Ø = 3mm). The target was located 95mm beneath the "skin" surface of the Phantom. It is important to notice that the Phantom's internal density and external texture resemble those observed in real human tissues. Though the 3D location of the target within the Phantom was identical for all trials, the location of the entry point was randomly varied across trials to avoid the production of stereotypic similar trajectories across trials. We compared the performances of individuals placed in two conditions of movement guidance: either they had to use a CAS intracorporeal guidance system providing the surgeon with 2D-visual information about the position of the needle with respect to the optimal trajectory; or, they had to carry out the same task guiding the puncture with the information furnished by the TDU device.

## 2.2. TDU guidance system

### 2.2.1. Material

For this experiment, electrotactile stimuli were delivered to the dorsum of the tongue via a ribbon TDU derived from the one developed by Bach-y-Rita. It consisted in an array of 36 electrotactile electrodes (6×6 matrix, radius: 0.7mm each), embedded in a 1.5×1.5 cm plastic strip (Fig. 1a). The tip of the TDU was inserted in the oral cavity and held lightly between the lips, maintaining the array in close and permanent contact with the surface of the tongue (Fig. 1b). A flexible cable, made of a thin (100 μm) strip of polyester material, connected the matrix to an external electronic device delivering the electrical signals activating the tactile receptors of the tongue. The frequency of the DC stimulating pulses was fixed to 50Hz across trials and subjects. Because of the conductive properties of the saliva, the TDU only requires a 5-15V input voltage and a 0.4-4.0mA current. Each gold-plated circular electrode (Ø = 1.4mm, inter-centre distance = 2.3mm) received monophasic pulses [17]. Furthermore, the literature evidences that the sensitivity on the tongue surface is not homogenous [15]. Bach-y-Rita proposed therefore to increase by a factor of 30% the stimulation in the posterior part of the matrix to compensate the low-threshold properties of this tongue region. Several trials preliminary to our experiment showed that some participants still couldn't feel the stimulation in the posterior part of the TDU matrix after some minutes of use. Therefore, it was arbitrary chosen to double the stimulation intensity of the 18 electrodes (the three posterior rows of the matrix) that correspond to the posterior part of the tongue.

-- **Insert figure 1 around here** --

*2.2.3. Principle of movement guidance under TDU*

The TDU system aims at transmitting sensory information to the tactile receptors of the tongue in order to provide surgeons with information permitting the directional control of his/her puncturing gesture. In order to compare the actual orientation of the ancillary with the optimal predefined trajectory, a simple and intuitive algorithm was chosen to code, onto the TDU, the orientation errors. The idea is that no electrical stimulation is provided if the needle is well oriented and is directed towards the target. Alternatively, if the tool is misoriented, an area of the TDU matrix is activated, thus indicating to the surgeon the incorrect orientation of the tool. Under this latter condition, the surgeon's task consists in moving the needle in a direction that will cancel the electrical stimulations. The information about the relative distance between needle's tip and target was delivered through modulation of the temporal frequency of sound beeps.

The activated zones on the matrix correspond to the direction of the inadequate needle tilt (and therefore a misoriented movement). Eight different orientations were defined: N, S, E, W, NE, NW, SE, and SW and a cross-stimulation indicated that the needle's tip overshot the target. This simple but sufficient principle of coding directional errors contributed to downsize the original Bach-Y-Rita's TDU (12×12 electrodes) and thus allowed the development of a smaller device (6×6 matrix of electrodes) easily and comfortably insertable in the oral cavity. Furthermore, it is to note that we simulated a "safety cylinder" ($\varnothing$ = 3mm) corresponding to the 3D extrusion of the target between the entry-point and the target. When the needle's tip remained within the limits of the cylinder, no activation was sent to the TDU, conversely, if it exited this cylinder, the directional-error coding principles applied. For instance, figure 2 schematises an example of the activation of the North-western area (NW) of the TDU resulting from an erroneous NW tilt of the needle (the tip being outside the "safety cylinder"). As figured by the green arrows, to eliminate the vibrational sensation perceived on the tongue, the surgeon has to move the needle in the South-eastern direction (SE). Moreover, in order to provide the surgeon with information about the extent the tool misorientation (distance to the planned trajectory), we chose to regulate an intensity of electro-stimulation that is proportional to the amplitude of the misorientation (i.e., the more the angle between the ancillary and the "ideal" planned trajectory augmented, the more the intensity of the stimulation increased).

-- **Insert figure 2 around here** --

2.3. Visual guidance system

For the experimental condition involving the visual guidance of the puncture we used a CAS system (Fig. 3). A dedicated software processed the coordinates of both the needle's tip and the target. The visual information available on a video screen (21 inches, Sony®) provided a parsimonious two-dimensional display of the needle's tip and the target relative positions. A grey immobile cross figured the direction of the optimal trajectory and a mobile red cross indicated the actual position of the needle's tip. This 2D view represented a plane centred on the target and perpendicular to the optimal predefined trajectory. Subjects had to maintain the needle within a zone of tolerance (equivalent to the one used in the TDU condition) while inserting the needle into the Phantom. Note that a 1:10 scale ratio between the visual representation and the actual displacements of the needle allowed increasing the accuracy of the motor corrections. Similarly to the TDU condition, information about the relative distance between needle's tip and target was delivered through modulation of the temporal frequency of sound beeps.

It is to underline that the only information conveyed by the system of visual guidance was reduced to a real-time display of the needle's orientation so that both systems (TDU and visual guidance systems) may be considered as affording the same parsimonious type of information.

-- **Insert figure 3 around here** --

2.4. Procedure

Fourteen subjects (18-35 years; 12 males and 2 females, students at Grenoble University) participated in this experiment. They were naive as to the goals of the experiment and had normal or corrected-to-normal vision. None of them reported suffering from troubles that could impair movement guidance.

Instead of testing all subjects for both conditions, it was decided to separate them in two groups assumed as independent (2 x 7 subjects). Indeed, it is well known that the learning effect that can be observed when a single group carries out two successive experiments (TDU after V or V after TDU), induces a much more important bias than the bias due to two groups assumed as independent (for this experiment, subjects had similar age and socio-cultural origins and were chosen randomly for each TDU or V condition). The first group (7 subjects) carried out one succession of 10 punctures using only the information provided by the TDU system whereas the second (7 subjects) accomplished the same task using the visual guidance system. Participants were instructed to reach the target at their best speed-accuracy trade off. No additional information about the spatial localization of the intracorporeal target or the adequate orientation of the needle was given. We excluded from the present analysis trials that lasted more than 200s (10% of a total of 140 trials).

In the TDU condition, participants were familiarized with the electrotactile information. This phase aimed at determining individuals' optimal intensity threshold for the lingual electrotactile stimulation and explaining the associations between corrections in the orientation of the needle and the information provided by the TDU. Subjects were thus enrolled in a training task in which they had to discriminate and associate stimulations provided by the TDU (among the 8 azimuths the matrix can code) with a specific needle orientation. The TDU guidance experiments only began once the subjects succeeded in this training task (among all subjects, the observed maximum training time was below 15 minutes).

2.5. Data acquisition and analyses

In order to capture the 3D position of the ancillary during the puncture, a Polaris® system (Northern Digital Inc.) was used. This optical localizer device is made of infrared (IR) reflectors secured on the needle and monitored by two infrared cameras. The system is then able to track (at a sampling frequency of 20Hz) the needle position relative to the reference frame of the surgical table. Thus, the displacement vector and rotation matrix of the needle is fully determined with a resolution of 0.2mm in all directions. In the two conditions (TDU and Visual guidances) three miniature IR reflectors were therefore secured on the needle and three additional IR reflectors were set on the Phantom so that to express the position of the needle with respect to the target within the artificial abdomen (Phantom frame of reference). Dedicated software compared in real time the needle's actual position with respect to an optimal trajectory defined as the straight line joining entry point to the target. The target and therefore the optimal trajectory were defined prior to movement through 2D ultrasound imaging of the Phantom. For this, the ultrasound (US) probe was equipped with IR reflectors. Each time an image was recorded, the six position parameters of the probe were also recorded, thus localizing the 2D US image in the 3D space.

Data about the 3D positions of the needle tip (see Figure 4) were processed under Matlab 6.5 (Mathworks®, USA). For each trial we measured 1) the sum of the point-to-point vectors in order to assess the length of movement amplitude (MA), 2) the maximal distance (MD) in calculating — at each sampling interval — the distance between the actual position of the needle tip and the predefined "optimal" trajectory and finally, 3) the total movement time of the puncture (TMT).

Analyses involve comparing data collected in V and TDU. As the sample size (n=7) only allows a non-parametric statistical test, three non-parametric analyses of variance (Friedman's ANOVA) were processed in order to assess the evolution of each parameter along the ten trials of a movement production. Paired Mann-Whitney U tests were finally used to test for differences between trials carried out under TDU and visual guidance systems.

-- **Insert figure 4 around here** --

## 3. Results

After a global analysis of the results provided by the TDU guidance experiments, in terms of target hit performance, the spatial-temporal parameters of movement in the TDU and V conditions are compared.

3.1. Overall performance under TDU guidance

While in the V condition 100% of the trials were successful, 96% of the trials under TDU guidance reached the target within 200s. Variable performances were observed across participants. Indeed most subjects reached the target for all trials (100% performance) and one subject achieved an incomplete performance (80% of successful trials). These results show that, after limited training (less than 15 minutes per subject), the guidance via lingual electrotactile stimulation appears to be efficient to control intracorporeal puncture gestures. We may suppose that the 100% baseline performance evidenced for every subject and every trial under visual guidance should be reached after sufficient training under the TDU.

3.2. Spatial-temporal Performances: TDU vs. Visual Guidance

3.2.1. *Movement Amplitude (MA)*

Figure 5 shows the mean MA (± standard deviations) for every trial for the TDU and V conditions. Statistical analyses evidence a significant decrease of MA as a function of the trial number (Friedman's ANOVA, Chi² = 17.1, $p < .05$) in the TDU condition and a stable performance in the V condition (Friedman's ANOVA, Chi² = 4.1, $p = $ ns). Moreover performances in both TDU and V conditions were significantly different for the first 6 trials (all $ps < .035$) and not for the last 4 trials (trials No. 7, 8, 9, 10, all $ps = $ ns).

-- **Insert figure 5 around here** --

3.2.2. *Maximal Distance (MD)*

Figure 6 plots the mean MD (± standard deviations) for every trial for the TDU and V conditions. Statistical analyses evidence a tendential decrease of MD as function of the trial number (Friedman's ANOVA, Chi² = 15.5, $p < .10$) in the TDU condition and a stable performance in the V condition (Friedman's ANOVA, Chi² = 2.2, $p = $ ns). Differently to the previous parameter (i.e., movement duration), performances in both TDU and V conditions were always significantly different (trials No. 1 to 10, all $ps < .025$), MD being greater in the TDU condition.

-- **Insert figure 6 around here** --

3.2.3. *Total Movement Time (TMT)*

Figure 7 shows the mean TMT (± standard deviations) for every trial for the TDU and V conditions. Statistical analyses evidence a significant decrease of TMT as function of the trial number (Friedman's ANOVA, Chi² = 16.8, $p < .05$) in the TDU condition as well as in the V condition (Friedman's ANOVA, Chi² = 14.3, $p < .05$). Moreover performances in both TDU and V conditions were not significantly different for most trials (trials No. 2 : 7 and 9 : 10, all $p$s = ns).

-- **Insert figure 7 around here** --

## 4. Discussion / Conclusion

The present exploratory research aimed at evaluating the possibility of movement control via a Tongue Display Unit in the context of guiding intracorporeal percutaneous punctures. Despite the low number of subjects, results reveal the usability of the TDU system in a surgical task and are very encouraging for further development of electrotactile movement guidance devices. On a general basis, results highlight the fact that electrotactile stimulation of the tongue to guide a movement appears to be efficient: Under TDU guidance, subjects indeed reached the intracorporeal target in 96% of the trials.

When comparing performances in conditions of visual and TDU movement guidance (V vs. TDU), it follows from the above analyses that though temporal data (TMT) fails showing a difference between the two conditions of movement guidance for most trials, both spatial parameters (MD & MA) tend to evidence a slight superiority of visual guidance. However, concerning MA, this superiority disappears after the 7$^{th}$ trial. Thus, only the MD parameter still shows differences between the two conditions of guidance. These limited differences between V and TDU conditions are somewhat surprising because there are evidences demonstrating that evaluation and reproduction of orientations is far more efficient in the visual than in the tactile modalities [18]. The present data are therefore very encouraging, assuming that a more significant training period should improve the TDU guidance performances, specifically as concerns MD.

Concerning the usability of the TDU device, numerous subjects noted that the system appeared more "intuitive" than the visual system. We may propose that the surprising and remarkable performance of operators under TDU guidance could be explained by the involvement of different processing of action-related information operating when correcting for a misorientation of the surgical ancillary. Indeed, whereas under visual guidance, movement corrections evidently involve the processing of spatial information (angular correction), under TDU guidance the same corrections might only involve a more primary behaviour relying on the avoidance of the tactile stimulation. Therefore, introducing the ancillary while at the same time avoiding the tactile stimulation insures the operator to reach the target safely. Such a difference in the processing of movement-related information may explain the rapid and almost complete adaptation of naïve operators to the TDU guidance of the puncture.

Nevertheless one may think that anatomical, physiological and ergonomic constraints can hamper the usability of the TDU device. On the anatomical and physiological side, the well-documented heterogeneous distribution of tactile receptors on the tongue surface (lower density in the posterior region) has to be taken into account in regulating the intensity of the stimulation. The present results show that doubling the intensity for Northern electrodes (N, NE & NW) was not sufficient enough. Indeed, it appears that 64% of the stimulations (summation across trials and subjects) were located in the posterior part of the TDU matrix (i.e., the Northern electrodes). Such an important proportion may signal that subjects experience difficulties in perceiving Northern stimulations and therefore made systematic orientation errors in this direction. It nevertheless appears that though the stimulating parameters used in the present exploratory study are not totally optimized, behavioural results show that the guidance of a highly accurate surgical gesture via the spatial information afforded by the TDU is possible. From now, further developments of TDU devices will have to take into account this important outcome. For instance, psychophysical studies of the heterogeneous tactile sensitivity of the tongue could help optimizing the electrotactile tongue stimulation.

Moreover, because of its 'wired' presentation the present TDU device is not usable in the context of real surgical practise. Indeed, using such a wired device in an operating room would constrain the movements of the surgeon and potentially have dangerous drawbacks. To counter this problem, we have developed with the support of the French company Coronis-Systems® a first wireless 6×6 TDU prototype (Fig. 8) inserted in a dental retainer including both microelectronics and power supply. This new TDU, once functionally validated, should be more easily accepted by surgeons than the wired device in the context of therapeutical procedures.

**-- Insert figure 8 around here --**

To conclude, the results of the described experiment have shown the feasibility of the TDU guidance. Indeed it has to be stressed that the tactile stimulations of the tongue seems to provide accurate and costless (on a cognitive viewpoint) orientation information that can be used with only little prior training. It may be therefore suggested that a TDU (dispensing orientation information via a non-visual channel) could be used in an attempt to liberate the surgeon's visual resources so that she/he could concentrate on the clinical aspects of her/his own gesture. This new device needs of course complementary studies to be more quantitatively validated, particularly with expert surgeons when using more complex trajectories.

**Acknowledgments**

The authors wish to thank Paul Bach-y-Rita for providing the 12×12 TDU device and for his advices and discussions on sensory substitution.

# Figures caption

**Figure 1:** Tongue Display Unit. a) The 6×6 array of electrodes embedded at the tip of the plastic trip and placed on the tongue. b) Subject using the TDU to guide a puncturing needle in an abdominal Phantom. The subject gazes at the needle inserted in the phantom.

**Figure 2:** Example of movement guidance under TDU (figure not to scale). Coding of the difference between the orientation of the ancillary (in red) and the "ideal" planned trajectory (blue line) at each moment in time during the puncture. As noted in the text, an inadequate orientation of the ancillary (e.g., a tilt in the NW direction) results in the activation of the three electrodes located in the NW corner of the TDU matrix (red cylinders). In this case, others unitary electrodes remain inactivated (small grey cylinder). As figured by the green arrows, the surgeon has to move the needle in the SE direction to eliminate the stimulation perceived on the tongue.

**Figure 3:** Visual guidance system. Left: the subject gazes at a video screen. Right: Schematic view of the screen display. This 2D visualization system figured the tip of the needle (red cross), the target (grey cross) and the orthogonal section of the "safety" cylinder (dotted circle).

**Figure 4:** Three-dimensional view of an exemplar trajectory (red line) and the virtual "safety cylinder" revolving around the ideal planned trajectory ($\varnothing = 3mm$, in transparent grey).

**Figure 5:** Evolution of movement amplitude (MA) as a function of the condition of movement guidance (TDU and V) and of the trial

**Figure 6**: Evolution of maximal distance (MD) as a function of the condition of movement guidance and of the trial.

**Figure 7:** Evolution of total movement time (TMT) as a function of the condition of movement guidance and of the trial.

**Figure 8:** Prototype of the 6×6 wireless TDU inserted in a palatal prosthesis

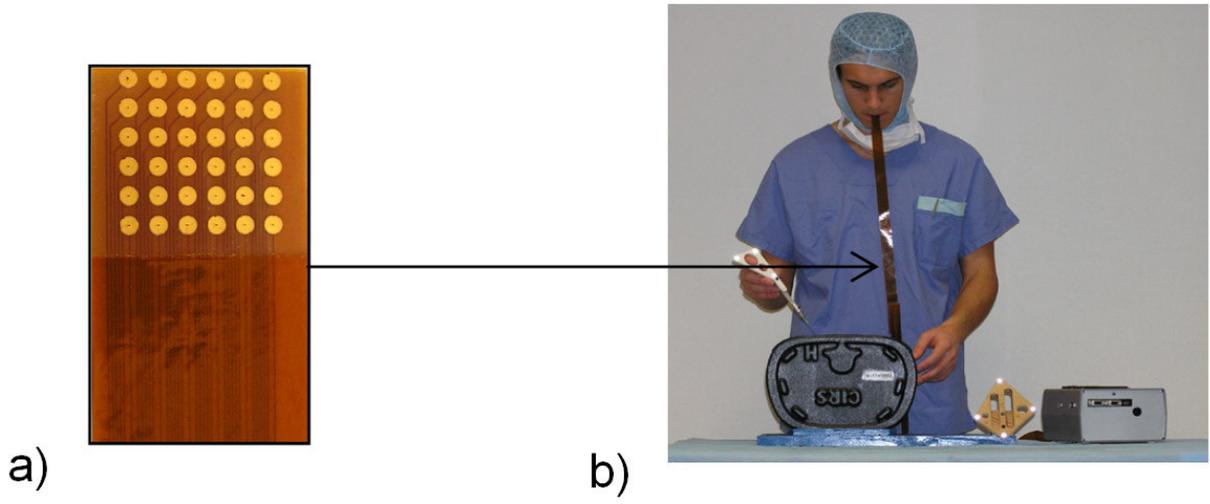

**Figure 1**

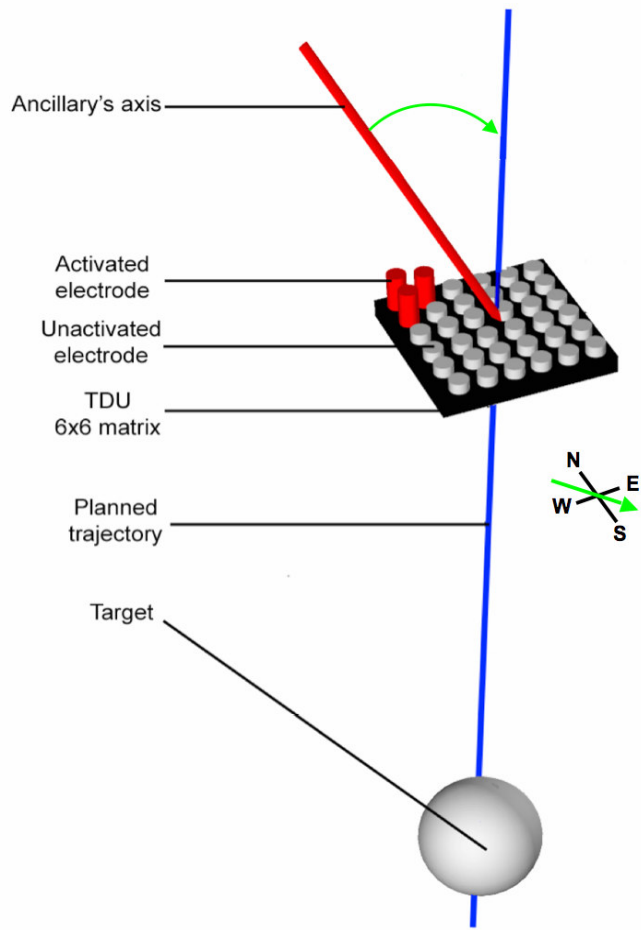

**Figure 2**

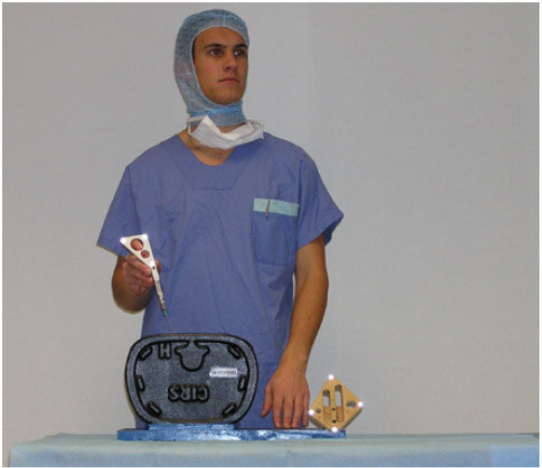 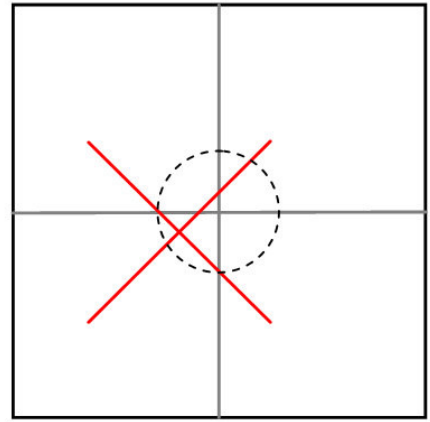

**Figure 3**

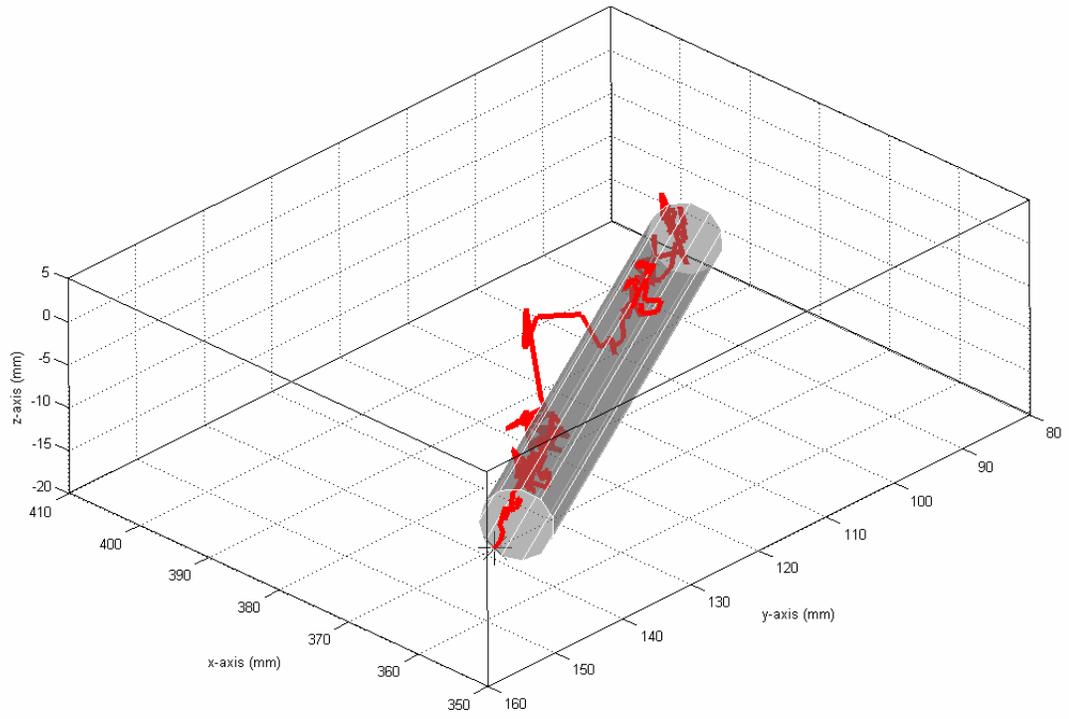

**Figure 4**

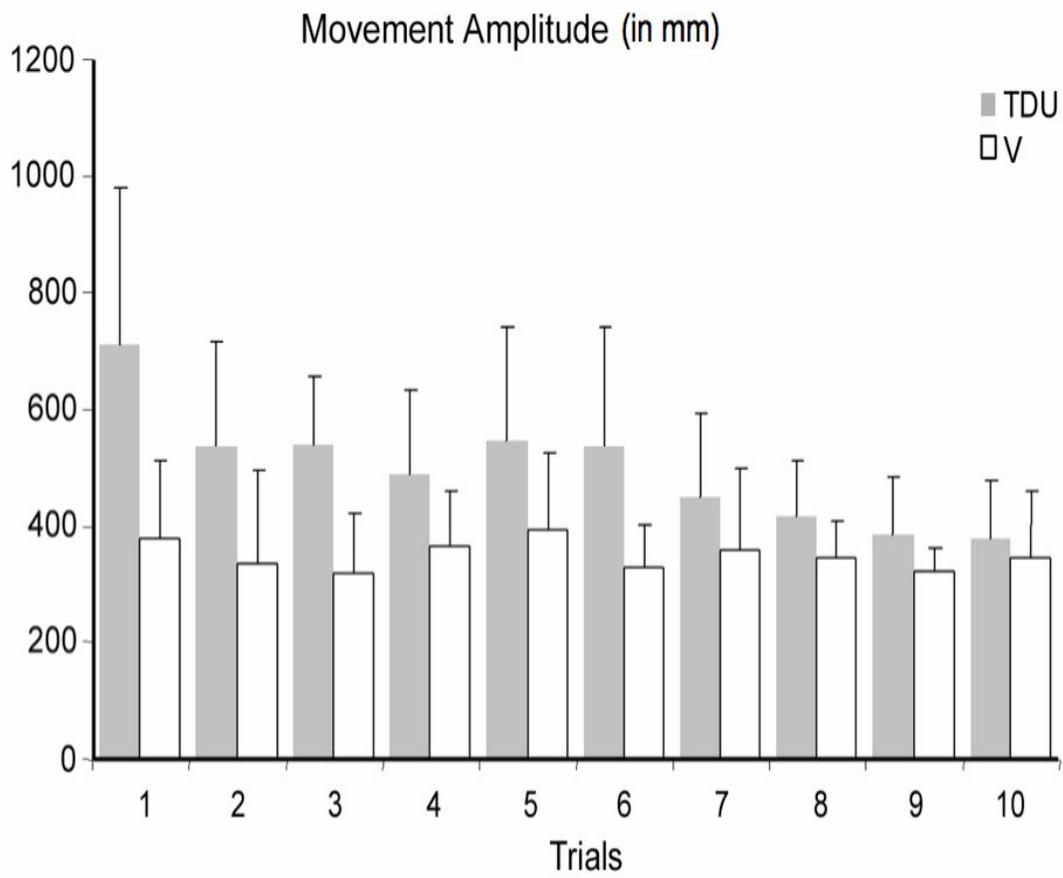

**Figure 5**

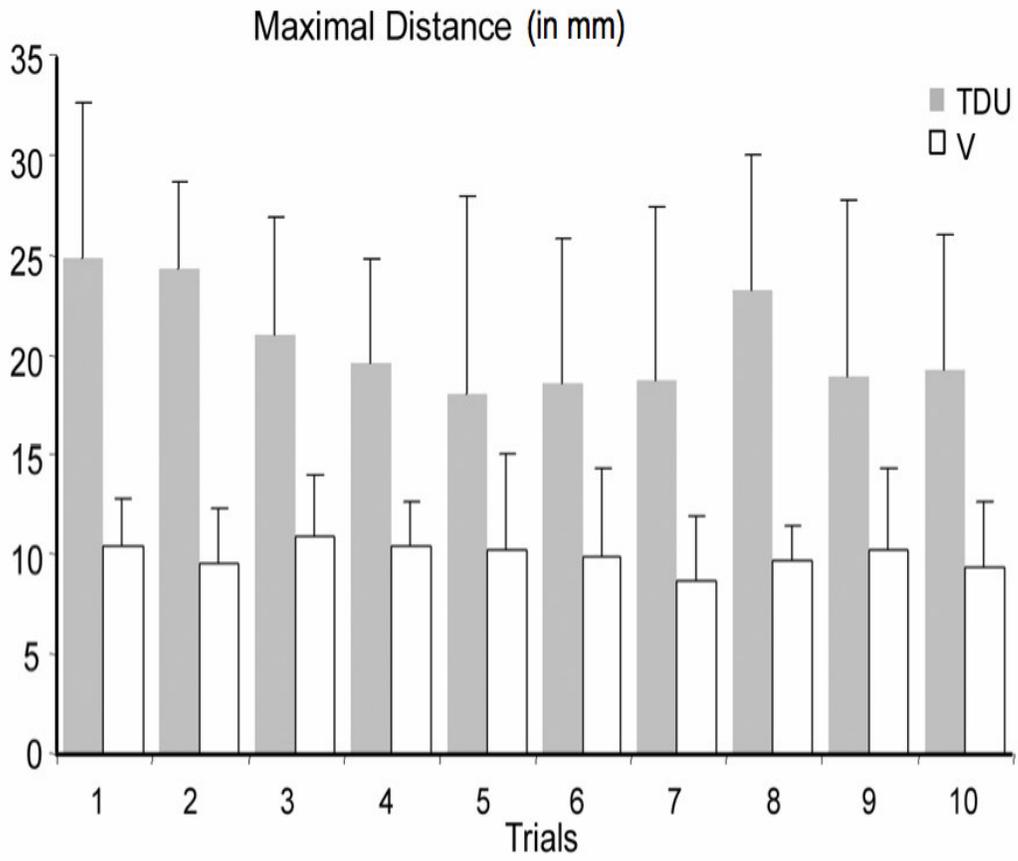

**Figure 6**

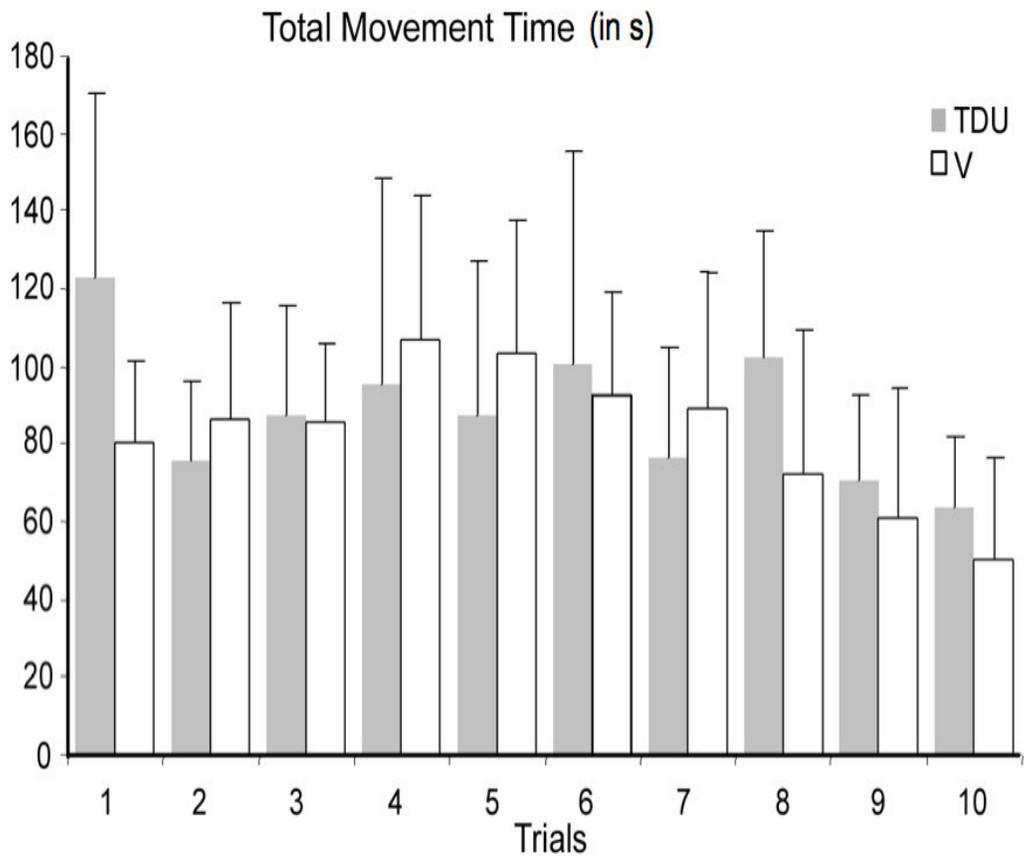

**Figure 7**

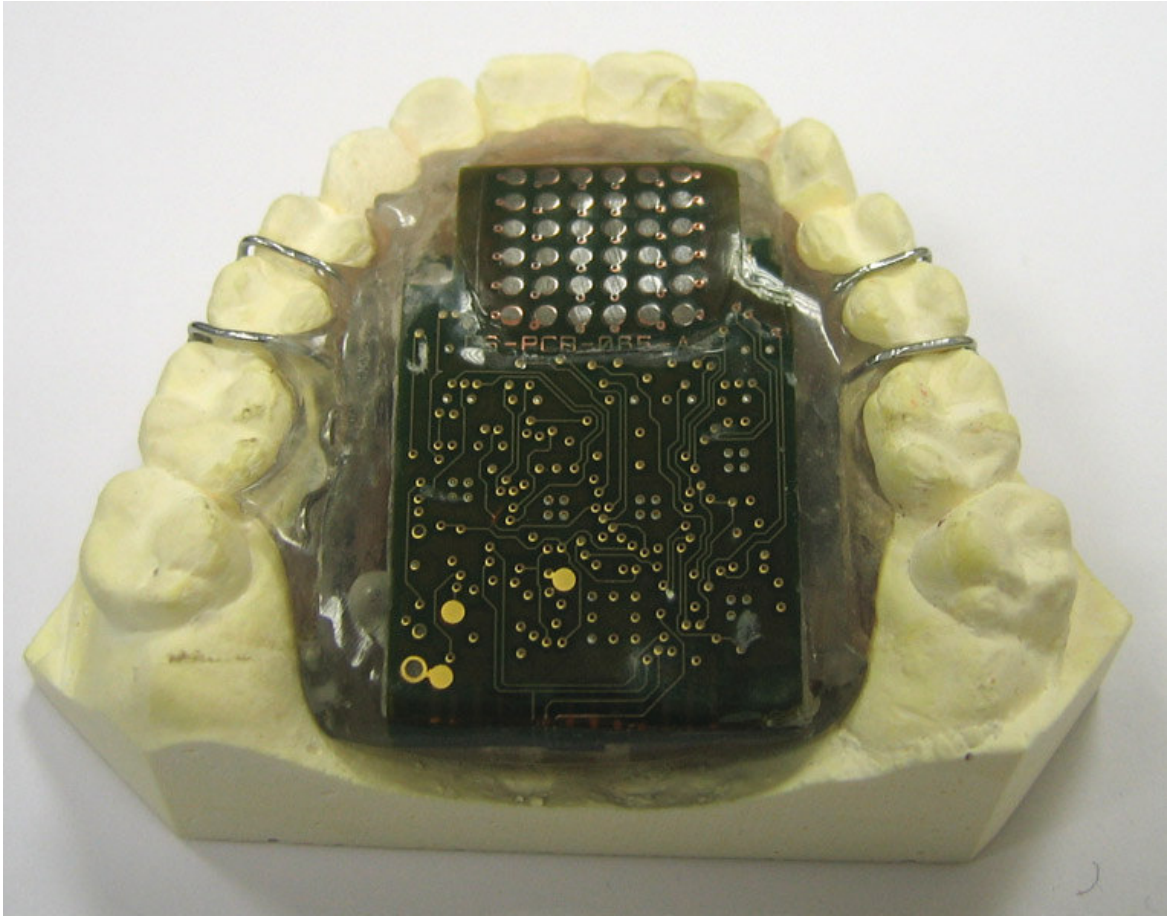

**Figure 8**